\documentclass[prb,twocolumn,floats,aps]{revtex4}
\usepackage{graphicx,graphics,color,epsfig} 
\usepackage{bm}
\usepackage{amssymb}
\usepackage{amsmath}
\usepackage{amsfonts}

\begin{document}

\title{Energy dependence of localization with interactions and disorder:  The generalized inverse participation ratio of an ensemble of two-site Anderson-Hubbard systems}
\author{J. Perera}
\author{R. Wortis}
\affiliation{Department of Physics \& Astronomy, Trent University,
1600 West Bank Dr., Peterborough ON, K9J 7B8, Canada}
\date{\today}

\begin{abstract}
After Anderson's prediction of disorder-induced insulating behavior, extensive work found no singularities in the density of states of localized systems.
However, Johri and Bhatt\cite{Johri2012a,Johri2012b} recently uncovered the existence of a non-analyticity in the density of states near the band edge of non-interacting systems with bounded disorder, in an energy range outside that captured by previous work.  
Moreover, this feature marks the boundary of an energy range in which the localization is sharply suppressed.
Given strong current interest in the effect of interactions on disordered systems, we explore here the effect of a Hubbard $U$ interaction on this behavior.
We find that in ensembles of small systems a cusp in the density of states persists and continues to be associated with a sharp suppression of the localization.
We explore the origins of this behavior and discuss its connection with many-body localization.
\end{abstract}
\maketitle

\section{Introduction}
\label{sec-intro}

\begin{figure*}
\includegraphics[height=1.6 in]{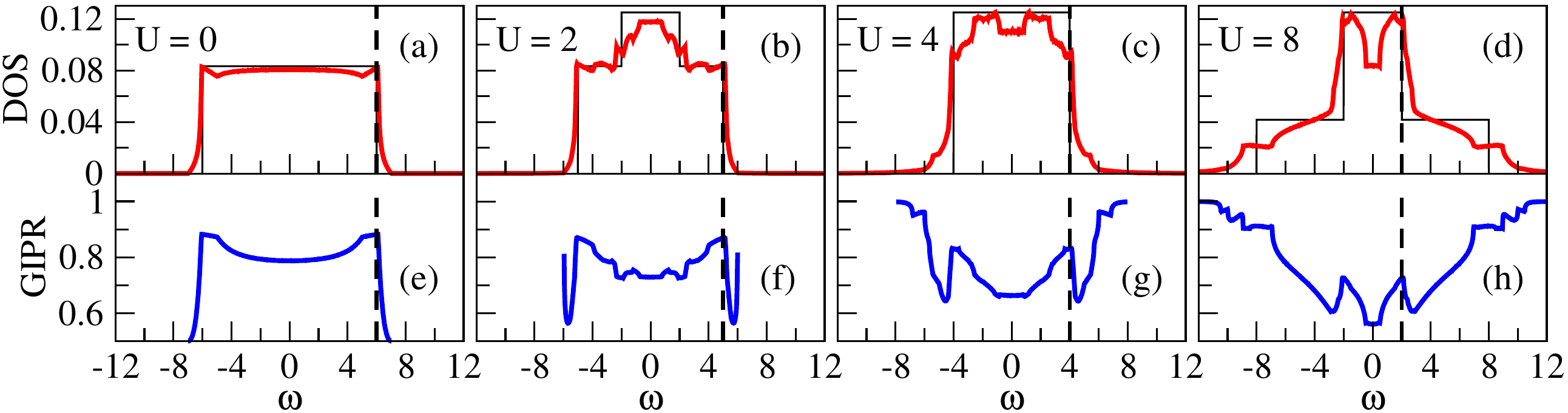}
\caption{\label{main}
(Color online)
Panels (a)-(d) show the ensemble-averaged density of states for $W/t=12$ and $U/t=0$, 2, 4 and 8 respectively.  The thin black lines indicate the atomic-limit density of states.
Panels (e)-(h) show the ensemble-averaged generalized inverse participation ratio at the same $U$ values.
The vertical dashed lines mark energies at which a cusp appears in the density of states and a sharp drop occurs in the generalized inverse participation ratio.
Ensembles consist of 100-150 million systems and the energy bin width is 0.019t-0.025t.
}
\end{figure*}

Unconventional superconductivity and other novel electronic behaviors in transition metal oxides continue to drive the study of strong electron correlations.
Doping which breaks translational invariance is generally required to obtain these novel behaviors, raising the question of how disorder affects strongly correlated systems.
Meanwhile, since the seminal work of Anderson in 1958,\cite{Anderson1958} localization of single-particle states by disorder has been the subject of intense study in a wide range of systems.\cite{50years}
Experiments have suggested that electron-electron interactions might drive a delocalization transition,\cite{Kravchenko2004} and there has been recent rapid progress in addressing the question of how interactions affect localized systems.\cite{Basko2006,Pal2010,Serbyn2013,Kjall2014,Huse2014}

In this paper we ask how electron-electron interactions affect several novel features recently pointed out in non-interacting disordered systems.\cite{Johri2012a,Johri2012b}
After Anderson first predicted localization, significant effort went toward establishing whether or not any feature in the density of states was associated with localization.  
Edwards and Thouless\cite{Edwards1971} showed that the density of states is analytic in a wide energy range around the middle of the band for all continuous distributions, and Wegner\cite{Wegner1981} showed that the density of states neither vanishes nor diverges anywhere inside the band for Gaussian distributed disorder. 
It therefore came as a surprise when Johri and Bhatt\cite{Johri2012a,Johri2012b} uncovered the existence of a non-analyticity in the density of states of systems with bounded disorder.  
Their result does not contradict earlier work.  Rather, the non-analyticity they found occurs outside the energy range addressed by Edwards and Thouless.  

Beyond the novelty of this singularity, Johri and Bhatt showed that a corresponding singularity exists in the inverse participation ratio, a measure of localization, and demonstrated that these singularities mark the boundary of a region at the edge of the band dominated by resonant states, also known as Lifshitz states.  
Moreover, while this behavior persists in large systems and in two and three dimensions,\cite{Johri2012a} Johri and Bhatt demonstrated that the key features can be seen and understood analytically in the simple case of an ensemble of two-site systems.\cite{Johri2012b}

Here we consider an ensemble of two-site Anderson-Hubbard systems and calculate both the density of states and a generalized form of the inverse participation ratio applicable to interacting systems.  
Fig. \ref{main} summarizes our key results.  
The presence of local interactions results in a richer energy dependence of both quantities.  
In particular, the minimum in the inverse participation ratio no longer occurs at the band edge but instead moves closer to the Fermi level.
We present below an exploration of how these changes arise, highlighting two distinct types of resonance which can occur in interacting systems and their role in our results.

\section{The model}

The Anderson-Hubbard model is a tight-binding Hamiltonian which includes nearest-neighbor hopping $t$, on-site Coulomb repulsion $U$, and site potentials chosen from a uniform distribution of width $W$.  
\begin{eqnarray}
\mathcal{H}
&=& 
t \sum_{\langle i,j \rangle,\sigma} {\hat c}_{i\sigma}^{\dag} {\hat c}_{j\sigma}
+ \sum_i U {\hat n}_{i\uparrow} {\hat n}_{i\downarrow} 
+ \sum_{i,\sigma} (\epsilon_i - \mu) {\hat n}_{i\sigma}
\end{eqnarray}
${\hat c}_{i\sigma}^{\dag}$ is the creation operator for an electron with spin $\sigma$ at lattice site $i$,
${\hat n}_{i\sigma}={\hat c}_{i\sigma}^{\dag} {\hat c}_{i\sigma}$, and   
$\langle i,j \rangle$ refers to nearest neighbor pairs.
The site potentials $\epsilon_i$ are chosen from the distribution $P(\epsilon_i)=\Theta(W/2 - |\epsilon_i|)/W$ where $\Theta$ is the Heaviside function.  
$\mu$ is the chemical potential.

We consider here ensembles of two-site systems so $i,j=1,2$, and we focus on the case of half filling so $\mu=U/2$.
For each system in an ensemble, the site potentials $\epsilon_1$ and $\epsilon_2$ are different but all are chosen from the same distribution.
Following Johri and Bhatt, it is convenient to use the coordinates
\begin{eqnarray}
x &=& {\epsilon_1+ \epsilon_2 \over \sqrt{2}} \ \ 
{\rm and} \ \ y \ = \ {\epsilon_1- \epsilon_2 \over \sqrt{2}}.
\end{eqnarray}
The phase space of all systems in an ensemble is represented by the diamond in the main panel of Fig. \ref{phase_space}.

\section{The density of states}

\begin{figure}
\includegraphics[width=\columnwidth]{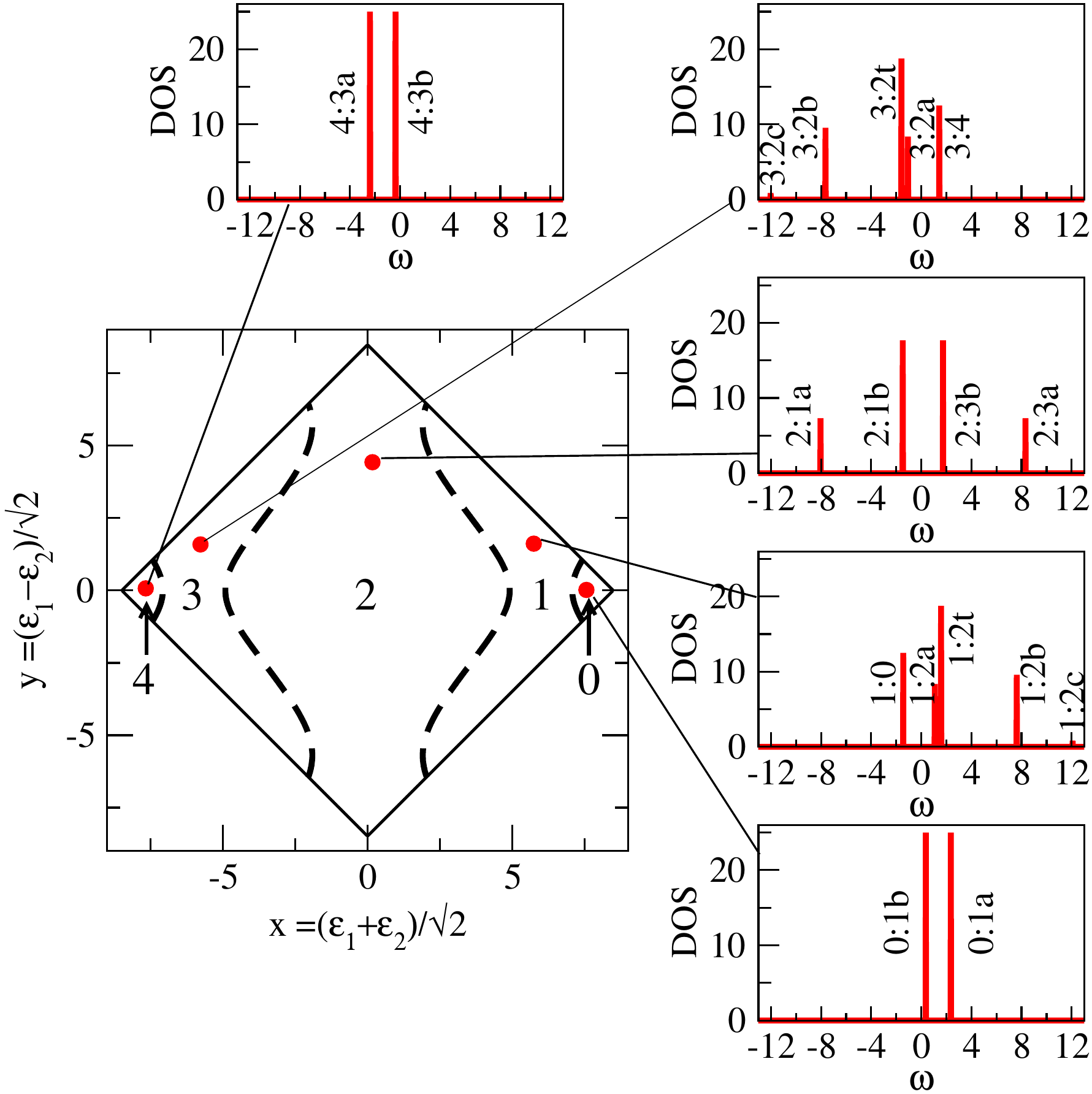}
\caption{\label{phase_space}
(Color online)
In the main panel, each point in the diamond specifies the site potentials for one 2-site system in an ensemble with disorder strength $W/t=12$.
Regions are labeled according to the number of particles in the ground state for interaction strength $U/t=8$.
Insets show typical single-system density of states plots for each region with the transition corresponding to each peak labeled. 
Energy bin width 0.02t
}
\end{figure}

The density of single-particle states for a single system is the average of the local density of states on the two sites calculated from the Lehmann representation of the retarded Green's function (see appendix).
\begin{eqnarray}
\rho(\omega) &=& {1 \over 2} \left( \rho_1(\omega) + \rho_2(\omega) \right) \\
{\rm where} \ \ \rho_i(\omega) &=& - {1 \over \pi} {\rm Im} \ G_{ii}^R (\omega) \label{ldos} \label{ldos}
\end{eqnarray}
The result consists of a small number of $\delta$-functions corresponding to the allowed single-particle transitions from the ground state.

The main panel of Fig. \ref{phase_space} shows the number of particles in the ground state in each region of the phase space,
and the insets show typical density of states plots in each of these regions.
For systems with a 0-particle ground state, there are two distinct peaks corresponding to transitions to the 1-particle bonding ($1b$) excited state or the 1-particle anti-bonding ($1a$) excited state.
For convenience we label the transition from the $n$-particle ground state to excited state $i$ the $n$:$i$ transition.
In this notation these two peaks correspond to the 0:1b and 0:1a transitions.
Each includes a spin-up and a spin-down contribution which, because our Hamiltonian is paramagnetic, occur at the same energy.
For systems with a 1-particle ground state, there are five distinct peaks corresponding to transitions to the 0-particle state and to six possible 2-particle states.  Three of the 2-particle states, the triplet states (2t), are degenerate.  The remaining 2-particle states (2a, 2b, and 2c) are linear combinations of the singlet state $|s\rangle = (|\uparrow \downarrow\rangle - |\downarrow \uparrow \rangle)/\sqrt{2}$ and the double-occupancy states $|2 0 \rangle$ and $|0 2\rangle$ in the Fock basis.
For systems with a 2-particle ground state, there are four distinct peaks corresponding to transitions to two 1-particle states (1a and 1b) and two 3-particle states (3a and 3b).
The transition options from the 3-particle ground state mirror those of the 1-particle ground state, and similarly the 4-particle options mirror those of the 0-particle ground state.

The density of states of an ensemble of two-site systems simply averages the contributions of all the systems.  
The density of states plots in Fig. \ref{main} (a)-(d) are histograms of the contributions in a sequence of frequency bins resulting in a smooth density of states for a large but finite number of systems.

In considering the evolution of the DOS with interaction strength, the atomic limit ($t=0$) provides a useful point of comparison.  This is indicated by the narrow black lines in panels (a)-(d) of Fig.\ \ref{main}.
Without interactions all sites are either empty or doubly occupied, and the atomic-limit DOS is identical to the distribution of site potentials: a plateau of height $1/W$.
On-site interaction results in singly-occupied sites which at $U=2$ create a narrow raised region at the band centre.
The width of the contribution from singly occupied sites grows with $U$, matching that from empty and doubly occupied sites at $U=4$ and exceeding it at $U=8$.

With hopping, the occupancy of each site is no longer integer, and there is an increase both in the number of distinct energies at which transitions can occur and in the range of weights associated with these transitions.
In the non-interacting case, this results in the two peaks at the band edge highlighted by Johri and Bhatt.\cite{Johri2012a,Johri2012b}
With interactions, the DOS gains even more structure.
The zero-bias anomaly in panel (d) has been explored in detail elsewhere.\cite{Wortis2010,Chen2010}

The particular feature on which we focus here is the persistence of sharp peaks, marked in Fig.\ \ref{main} by vertical dashed lines.
These peaks closely follow the edge of the contributions from empty and doubly occupied sites in the atomic limit,
which might suggest that the cusps would be lost when $U>W$ and all sites are singly occupied. 
In fact, when $U>W$ the shoulders seen around $\omega/t=\pm9$ in panel (d) 
evolve into sharp cusps, such that the DOS of the Mott insulator (not shown) looks like two copies of the $U=0$ case above.

\begin{figure}
\includegraphics[width=\columnwidth]{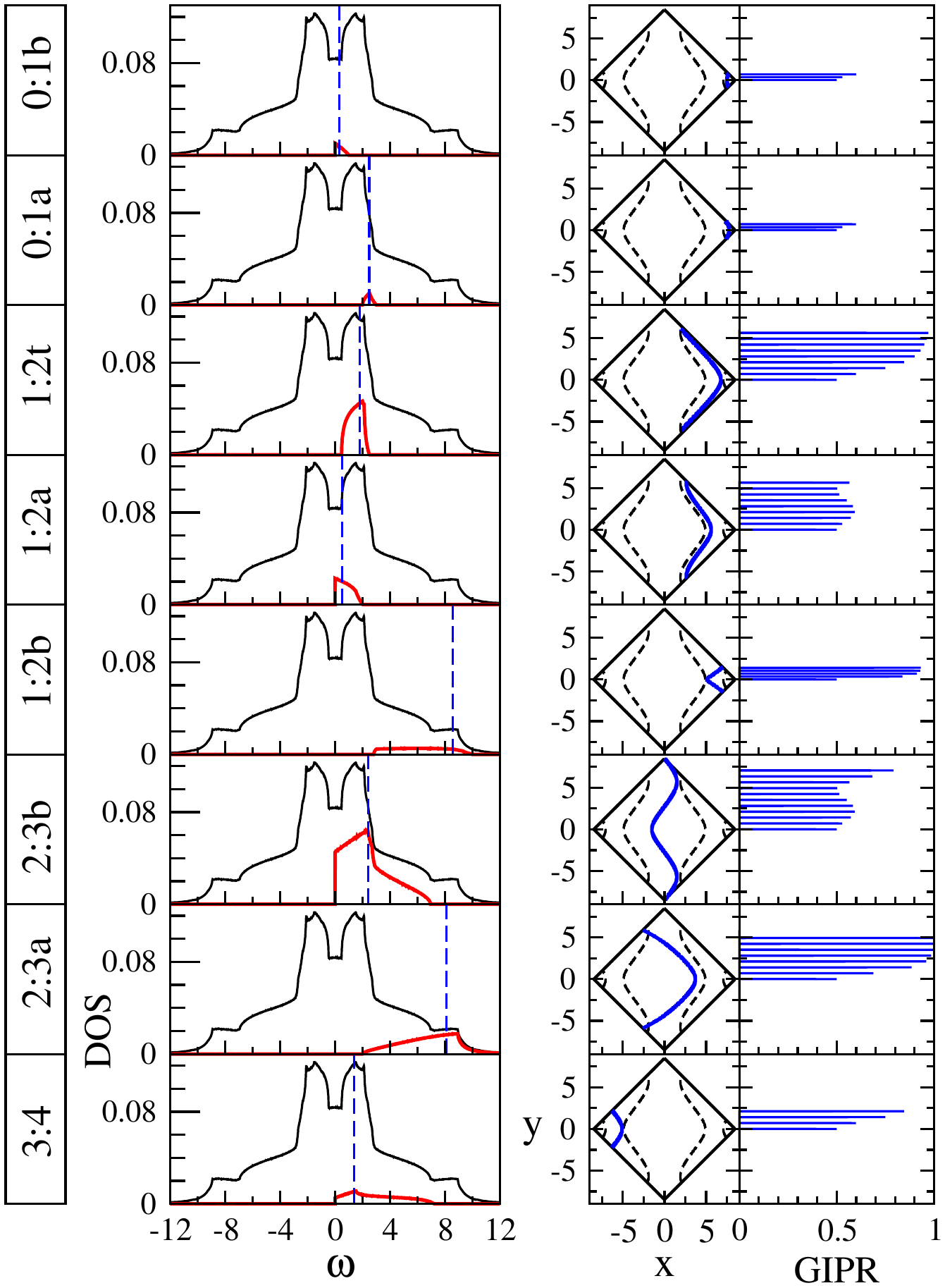}
\caption{\label{table}
(Color online)
Transition specific information for an ensemble of two-site systems with $W/t=12$ and $U/t=8$.
Each row corresponds to a different transition as labeled in column 1.
Only transitions which contribute above the Fermi level are shown.  
The 1:2c transition is not shown because its contribution is negligible.
Column 2 shows the contribution to the density of states made by this transition alone (red) as well as the full density of states (black) for comparison.
Column 3 shows the constant energy curve (blue) in phase space at the energy marked with a dashed line in column 2.  
Column 4 shows the generalized inverse participation ratio for this transition at a set of points along the constant energy curve shown in column 3.
}
\end{figure}

To provide further insight into the origin of these peaks, Fig.\ \ref{table} shows the DOS contributions from individual transitions.  
The first column indicates the transition, and the second shows the DOS contribution.
In the third column the phase space is shown with a blue curve which indicates all systems for which the specified transition occurs at the energy marked with a vertical dashed line in column 2. 
As the energy $E$ is increased, the shape of this curve for a given transition does not change but the curve shifts to the right.  
In the non-interacting case, the magnitude of the DOS contribution from each transition in each system is the same.  
Therefore the overall magnitude of the DOS at a given energy is simply proportional to the length of the constant energy curve at that energy.  
In the interacting case, there is variation in the magnitudes of the DOS contributions from different transitions, meaning that the line length alone is not the sole determinant of the DOS magnitude at each energy.  However, the main qualitative features of the DOS contribution from each transition are still reflected in the changes in line length.  

Consider, for example, the 1:2a transition.  The constant energy curve has the same shape as the boundary between the 2-particle and the 1-particle ground state regions.  
At $E=0$, the constant energy curve coincides with this boundary, and the transition makes its maximum contribution to the DOS.  As the energy is increased, the constant energy curve moves to the right declining in length roughly linearly with energy.
In this way the origin of the energy dependence of the DOS contributions of each transition can be seen, and from these the structure of the DOS as a whole.

In particular, what do we learn from Fig.\ \ref{table} regarding the origin of the DOS cusps?
As in the non-interacting case,\cite{Johri2012b} the peaks appear where a constant energy curve sits nearly tangent to the edges of the phase-space diamond, maximizing its length.  (See the 1:2t transition in Fig.\ \ref{table}.)
Unlike the non-interacting case, many transitions contribute at the same energy, resulting in variation in the shape of the peaks.

\section{The generalized inverse participation ratio}

To explore whether there is a connection between the structure in the density of states and variations  in the strength of localization, we have also studied a generalization of the inverse participation ratio, a standard measure of localization.  
In non-interacting systems, the inverse participation ratio at an energy $E$ is proportional to one over the size of a single-particle state with energy $E$.  
In interacting systems, single-particle states are not well defined.
Nonetheless, the following generalized inverse participation ratio (GIPR) can be defined:
\begin{eqnarray}
I(\omega) &=& {\sum_i \rho_i^2(\omega) \over [ \sum_i \rho_i(\omega) ]^2} \label{gipr}
\end{eqnarray}

$\rho_i(\omega)$, the local density of states as defined above in Eqn. (\ref{ldos}), has the form of a sum of weighted $\delta$-functions:  $\rho_i(\omega)=\sum_t w_{ti} \delta(\omega-E_t)$ where $E_t$ are the energies of single-particle transitions available from the ground state.
In many numerical calculations only the local density of states is known and only up to some nonzero energy resolution.
In such cases, when interactions are set to zero the GIPR has the same scaling behavior as the IPR in the limits of zero and infinite disorder, but the correspondence for intermediate disorder is less clear.\cite{Murphy2011}
In our case, however, we know the many-body eigenstates and can therefore distinguish the local-density-of-states contributions of individual transitions no matter how close they may be in energy.  
Regularizing the $\delta$-functions in the LDOS would result in a loss of information.
To avoid this we interpret Eqn. (\ref{gipr}) as
\begin{eqnarray}
I(\omega)={ \sum_i w_{ti}^2 \over \left[ \sum_i w_{ti} \right]^2} \ \ {\rm for} \ \ \omega=E_t
\end{eqnarray}
and zero otherwise.
In this way, in the absence of interactions the GIPR reduces to the IPR, expressed in terms of the single-particle eigenstate $\psi$ as $\sum_i|\psi_i|^4/\left[\sum_i|\psi_i|^2 \right]^2$.

Note that in an interacting system, the GIPR is associated with the size of a {\em transition} between two many-body states.
Column 4 of Fig.\ \ref{table} shows the GIPR values for specific transitions in selected individual systems.
For example, consider the 2:3a transition.  The blue line in column 3 shows all systems for which the energy of this transition is $8.1t$.  
The system at $y=5$ has a GIPR value of 1 for this transition.  
In this system, the LDOS associated with this transition is large on one site and close to zero on the other.  This is a transition which is localized primarily on a single site.
Alternatively, the system at $y=0$ has a GIPR of 0.5 for this transition.
In this case, the LDOS associate with the transition is roughly equal on the two sites, and the transition is as extended as it can be in a two-site system.

Column 4 of Fig.\ \ref{table} shows GIPR values for each transition for a selection of systems at different values of $y$ along the constant energy curve shown in column 3.
The GIPR values are independent of $x$.
For most transitions the GIPR value increases as $y$ increases.
For example, 2:3a transitions have GIPR values of 0.5 at $y=0$ and these values increase rapidly to 1 at higher values of $y$.

The 2:3b transition is particularly interesting:  There are two minima in the GIPR, one at $y=0$ and one at $y=\pm U/\sqrt{2}$.  
These two values represent the two distinct types of resonance which can occur in a strongly correlated system.  The $y=0$ resonance is the one present in non-interacting systems:  The potentials on the two sites are the same.  The $y=U/\sqrt{2}$ resonance is unique to strongly correlated systems:  Here the potential on one site is greater by $U$ than the potential on the other site.

From column 4 of Fig.\ \ref{table} it is clear that the transitions contributing to the DOS at any given energy may have a wide range of GIPR values, both because of contributions at the same energy from different transitions and because of variation in the GIPR within a single transition.
To obtain a single value at each energy we average over all the contributions.
In the non-interacting case studied by Johri \& Bhatt\cite{Johri2012a,Johri2012b}, the magnitude of the DOS contribution from each transition is the same, so the IPR values can be directly averaged.
In the interacting case we study here, there is wide variation in the magnitude of the DOS contributions.
We therefore average the GIPR values weighted by the magnitude of the corresponding DOS contributions.
Consider a transition $t$ in system $s$ at frequency $\omega_{st}$.
The DOS contribution for this transition is $\rho_{st}$ and the GIPR value is $G_{st}$.
We define the ensemble-average GIPR value 
\begin{eqnarray}
\langle G (\omega) \rangle &=&
{ \sum_s \sum_t G_{st} (\omega) \rho_{st} (\omega) \delta(\omega-\omega_{st}) \over \sum_s \sum_t \rho_{st} (\omega) \delta(\omega-\omega_{st})}
\end{eqnarray}

Fig.\ \ref{main} (e)-(h) show the ensemble-average GIPR versus energy for four values of $U$.  
At $U=0$ there is a gentle upward curvature in the centre of the band mirroring the gentle downward curvature in the DOS, consistent with the usual picture that localization is stronger when the DOS is smaller.  
Then, after reaching a sharp maximum there is an abrupt drop in the IPR at the edges of the band.
This drop has been shown\cite{Johri2012b} to come specifically from systems with extreme $x$ values but $y$ near zero.
The states in these systems have been referred to as resonant or Lifshitz states.  

When $U$ is nonzero, the GIPR, like the DOS, has much more structure.  
One pattern is an overall lowering of the GIPR over most of the width of the band as $U$ is increased.
This is consistent with weak interactions providing screening which reduces localization. 

A second pattern is the persistence of region of sharply suppressed localization. 
In particular this region moves towards the band center as $U$ increases.
What causes this abrupt drop?
Changes in the GIPR can occur at frequencies at which a particular transition starts or stops making DOS contributions or due to variations in the GIPR value within a specific transition.
The latter effect is generally smoother and smaller in magnitude.
The sharp drops in GIPR which move toward the center of the band with increasing $U$ are due to equal-site-potential resonances, as in the noninteracting case. 
There do not appear to be such dramatic features associated specifically with the $y=U/\sqrt{2}$ resonances unique to strongly correlated systems.

\section{Discussion}

In non-interacting systems, sharp peaks in the DOS were found to coincide with peaks in the IPR at the edge of an energy range dominated by resonant states.  
Focusing on ensembles of two-site systems at half filling, 
we find that in interacting systems the same pattern persists.
With available transitions dependent on the ground state of the system, both the DOS and the GIPR have more structure than in the non-interacting case.
Nonetheless, sharp peaks in the DOS continue to be associated with abrupt drops in the GIPR.

In non-interacting systems the decline in the IPR comes from resonant states in systems in which the potentials on both sites are similar.  
In strongly interacting systems, a distinction can be drawn between this matching of site potentials and a strongly-correlated resonance in which the site potentials differ by the onsite interaction strength $U$.  
Both types of resonance can be seen in the GIPR spectra of individual systems.
However, no feature in the ensemble average results can be uniquely associated with the strongly-correlated resonance.

Our results show a reduction of localization as interactions are turned on.
When the interaction strength becomes greater than the disorder and a Mott gap opens, the trend reverses, consistent with other work on interactions in disordered systems.\cite{Byczuk2005,Song2008,Henseler2008}
It is of particular interest that interactions shift the regions of suppressed localization toward the Fermi level, making this behavior more accessible to experiments.  
Both the DOS peaks and the GIPR dips should be observable in quantum dot systems.\cite{Wortis2015}

There has been recent rapid progress in understanding the phenomenon of many-body localization, focusing mostly on spin systems.  
It can be shown that the Hamiltonian of any system with a Hilbert space of dimension $2^N$ can be expressed in terms of spinors.\cite{Lychkovskiy2013,Huse2014} 
In this spinor picture, eigenstates differ in the flipping of individual spinors.  
Therefore, the GIPR presented here, which is a measure of the localization of transitions between many-body eigenstates, is providing information on the localization of these spinors.  

Moving away from half filling and exploring larger systems will be important directions for further study.




\section*{Acknowledgments}
We thank M.\ Kennett for a careful reading and W.A.\ Atkinson for helpful discussions, and we acknowledge support by the National Science and Engineering Research Council (NSERC) of Canada.

\appendix*

\section{The Green's function}

\begin{eqnarray}
G_{ii}^R(\omega) &=& {1 \over 2} \left( G_{ii \uparrow \uparrow}^R(\omega) + G_{ii\downarrow \downarrow}^R(\omega) \right) 
\end{eqnarray}
where 
\begin{eqnarray}
G_{ii\alpha\alpha}^R(\omega) &=& \sum_n \biggl\{
{|\langle \psi_n | {\hat c}_{i\alpha}^{\dag} |\psi_0\rangle|^2 \over \omega- (\Omega_n-\Omega_0) + i \eta} \nonumber \\
& & \hskip 0.3 in
+ {|\langle \psi_n | {\hat c}_{i\alpha} |\psi_0\rangle|^2 \over \omega + (\Omega_n-\Omega_0) + i \eta}
\biggr\}
\end{eqnarray}
Because the Hamiltonian conserves spin, $G_{ii\alpha\beta}^R(\omega)$ is zero for $\alpha \ne \beta$.
$\psi_0$ is the many-body eigenstate with the lowest grand potential $\Omega_0=E_0-\mu N_0$ where $E_0$ and $N_0$ are the energy and the particle number corresponding to this state. $\psi_n$ are all other many-body eigenstates with grand potentials $\Omega_n$.
In non-interacting systems the distribution of this local Green's function is an order parameter for the Anderson transition,\cite{Mirlin1994,Schubert2010} and it is also being explored as a measure of the many-body localization transition.\cite{Nandkishore2014}


\end{document}